\documentstyle[preprint,aps]{revtex}
\begin{document}
\title{Models of Fractal River Basins}
\author{Marek Cieplak$^{(1)}$,
Achille Giacometti$^{(2)}$, 
Amos Maritan$^{(3)}$, Andrea Rinaldo$^{(4)}$, 
Ignacio Rodriguez-Iturbe$^{(5)}$ and Jayanth R. Banavar$^{(6)}$ }

\vskip 3.0cm

\address{$^{(1)}$ Polish Academy of Science, 02-668 Warsaw, Poland}
\address{$^{(2)}$ 
INFM Unit\'a di Venezia, Dipartimento
di Scienze Ambientali, \\ 
Calle Larga Santa Marta, I-30123 Venezia-Italy}
\address{$^{(3)}$ INFM Trieste and International School 
for Advanced Studies (SISSA), \\ 
I-34014 Grignano di Trieste, Italy}
\address{$^{(4)}$ Istituto di Idraulica "G.Poleni", Universit\`a di Padova, 
I-35131 Padova, Italy }
\address{$^{(5)}$ Dept. of Civil Engineering, Texas A\&M University,\\
College Station, TX  77843 } 
\address{$^{(6)}$  Department of Physics and Center for Materials Physics\\
The Pennsylvania State University\\
104 Davey Laboratory University Park, PA  16802 USA}
\vskip 0.3cm
\date{\today}
\maketitle
\begin{abstract}
Two distinct models for self-similar and self-affine river basins 
are numerically investigated. 
They yield  fractal aggregation patterns following non-trivial
power laws in experimentally relevant distributions.  
Previous numerical estimates on the critical exponents, when
existing, are confirmed and superseded.
A physical motivation for both models in the present framework
is also discussed.
\vskip 1.0cm

\end{abstract}

\newpage


\section{Introduction}
Experimental analyses of river networks 
\cite{Iri1} have shown clear examples of behavior analogous to 
critical phenomena characterized by the absence of a single well-defined 
length scale reflected in a power-law behavior of various quantities.
A fundamental question that arises from these observations is 
whether, in analogy with conventional critical phenomena, one may 
fruitfully classify this behavior into universality classes
that are characterized by different sets of exponents.  A related point 
is whether there exist scaling relationships between the various 
exponents of a given universality class.  Another vital issue
is the elucidation of simple models amenable to  analysis that
nevertheless capture some of the features of fractal fluvial patterns.

Within this framework a theoretical description of
the system needs to address two basic issues 
\cite{stark,manna0,OCN1,science,nagel,sun et al,Maritan et al,Banavar et al}.
First, a careful characterization of the
{\it topological} properties of the networks is essential
for understanding the basic transport mechanism in the basin.
References \cite{stark,manna0} are recent attempts in this direction.
Optimization principles have been exploited both numerically
\cite{OCN1,sun et al} and analytically \cite{science} to
explain the tendency of natural drainage networks to
evolve toward an optimal stable topology. General scaling
arguments can be found in \cite{Maritan et al}.

Second, a study of the dynamical evolution of the landscape (on
geological time scales) as a result of  interaction with external agents
(rain, wind etc) would be desirable \cite{nagel,Banavar et al}.

The present work will address only the first point. We will argue that,
despite much progress in the past few years, the
problem is not yet fully understood and deserves further analysis.
To this aim we will discuss,
based on physical arguments, two toy models of river networks.
The first model leads to  a {\it self-similar} river basin, and is
relevant when the erosional properties of the surface soil are strongly
heterogeneous. The second model considers the homogeneous basin
case and results in a self-affine river network.

Although the overwhelming majority
of the observational data are consistent with a {\it self-affine} description
(i.e. networks display a privileged direction), the marked
self-similarity of the basins with their own sub-basins
suggests a crossover from a {\it self-similar}
character above some length scale. This is one of the reasons for considering
models with both characters. It should also be emphasized that although some
of the features of the models presented here were previously discussed
in the literature (see below), we believe that both the physical
motivations and the analysis carried out here are essentially new.

The plan of the paper is as follows. In the next section few definitions
and scaling relations will be recalled. In Sec. III results for self-similar
are presented. Sec. IV  is dedicated to the self-affine counterpart.
In Sec. V few relevant experimental results will be briefly
reminded for the sake of completeness. Sec. VI will summarize our
findings along with some future perspectives.

\section{Definitions and scaling laws}
We define a river network as a spanning (loopless) tree on a 
lattice 
of linear size $L$ \cite{manna1}. Each site has exactly 
one output bond to one of its neighbors and no restriction on the number
of input bonds (three at most on a square lattice).
In a river basin, the area at any site is
defined  as
the number of sites upstream of the site 
connected by the network. From the computational point of view, 
it can also be regarded as a measure of the flow rate if a unit weight 
is assigned to each source thus simulating a unit constant precipitation.  

The equation for $s_i$, the area at a given site $i$, is 
\begin{eqnarray} 
\label{flow}
s_i &=& \sum_{j \in nn(i)} w_{i,j} s_j  \ + \ 1
\end{eqnarray}
where  $w_{i,j}$ is $1$ if $i$  
collects water from its nearest-neighbor ($nn$) site $j$ and $0$ otherwise. 

It is experimentally observed \cite{Iri1} and theoretical explained
\cite{sun et al,Maritan et al} that in river basins the
probability density $p(s,L)$ of a site having area $s$ in a system
of size $L$, has the scaling form
\begin{eqnarray} \label{pa}
p(s,L) = s^{- \tau} F(\frac{s}{L^{\phi}})
\end{eqnarray}
where $F(x)$ is a scaling function which takes into account 
finite size effects and $\phi$ is the finite size exponent. 

Similarly the distribution of {\it upstream} lengths has been also
predicted \cite{Maritan et al} and confirmed by field observation
\cite{rigon et al} to display the universal form:

\begin{eqnarray} \label{pl}
\pi(l,L) = l^{-\gamma} f(\frac{l}{L^{d_l}})  .
\end{eqnarray}

where $f(x)$ is the analogue of $F(x)$ and $d_l$ coincides with
the stream fractal dimension.
The upstream length is defined as follows. At a given site the areas
(see eqn. (\ref{flow})) of the nearest-neighbours
are checked. The site with the largest value leads to
the outlet. The site with the next-largest value is defined to be an upstream
site -- it indicates the longest path towards the source.
If two (or more) equal areas are encountered, one is randomly selected.
Alternatively, a burning algorithm \cite{manna0} could be also
employed.

In natural basins, the drainage area $s$ and
the stream length $l$ are related by  Hack's law \cite{hack}
\begin{eqnarray} \label{hack}
s \sim  l^{1/h}
\end{eqnarray}

The sub-basin from any site 
defined as all the upstream sites connected to it is characterized by 
typical longitudinal and transverse lengths $\xi _{\|}$ and $\xi 
_{\bot}$. 
For self-affine river 
networks one defines the Hurst (or wandering) exponent as
$\xi _{\bot} \sim \xi _{\|}^{H}$ with $H\leq 1$.  
Note that for self-similar river networks
(in which each rivulet originating 
from any site and proceeding to the global outlet is a fractal 
characterized by the same fractal dimension $d_{l}$), the Hurst exponent $H=1$.  

As one might expect the exponents are not independent.
For {\it self-similar} networks ($d_l>1$,$H=1$) $d_l$ determines all
the other exponents \cite{Maritan et al},\cite{sun et al}
\begin{eqnarray} \label{ss}
\phi=2 \;\;,\;\; h=d_l/2 \;\;,\;\; \tau=2-2/d_l \;\;,\;\;
\gamma=2/d_l
\end{eqnarray}
For {\it self-affine} networks ($d_l=1$,$H<1$) it is $H$ which defines all
the other exponents \cite{science}
\begin{eqnarray} \label{sa}
\phi=1+H \;\;,\;\; h=1/(1+H) \;\;,\;\; \tau=\frac{1+2 H}{1+H} \;\;,\;\;
\gamma=1+H
\end{eqnarray}

Two features of the above relations are worth mentioning. First there
is experimental evidence in the observed data that $H<1$ and $d_l>1$.
This apparent contradiction might be explained with the crossover
between the two regimes occurring at some length scale, as mentioned in the
introduction. Secondly it turns out from (\ref{ss}) and (\ref{sa}) that
{\it identical} values of the exponents are obtained from both cases
if $d_l=2/(1+H)$. This means that knowledge of the exponents
other than $d_l$ and $H$ cannot discriminate the self-similar
or self-affine character of the basin. In this respect a direct
measure of $d_l$ and $H$ appears to be crucial for its
characterization.
 
\section{Self-similar river network model}

We first discuss the model of \underline{self-similar} river 
networks.  Consider a network 
which is a square lattice of size $L \times L$ where the links of the 
rivers are identified with the bonds of the lattice.  Periodic boundary 
conditions are assumed in the left-right direction.  The bottom side of 
the square is defined to be the (fixed) outlet which collects the water 
that is flowing out.  Independent random numbers in the range (0,1) are 
assigned to the different bonds representing the erodability  
$P_{i}'s$, of the surface soil of the bond i.

The physical situation we have in mind leads to river network
formation based on an 
invasion percolation like mechanism \cite{chandler et al}.
The weakest erodable link is selected and assumed to 
be a part of the network. The second-ranking weakest link is then selected 
and so on. 
The process is iterated in the ensemble
of the remaining links until all sites are connected, i.e. 
they all have a route to the outlet.
Loops are excluded since
once  a preferred  route is selected, alternative routes formed due to the
presence of a loop would be energetically unfavourable.
Operationally, one thus obtains the network by incorporating
the regions in order of increasing strength so that no loops are formed,
yet all sites on the lattice are connected to the outlet sites. 
A variant of the above procedure leading to the same structure,
consists of starting from the links connected to the outlets,
selecting  the weakest one and proceeding invasively (i.e. always choosing
the new weakest link) in the new ensemble
of the interfacial links.
This model, which was originally introduced by Stark \cite{stark}
was subsequently rediscovered by Manna and Subramanian \cite{manna0}.
This is a model of headward growth of
streams away from a rift,  the weakest bond corresponding to the point most
susceptible to bank failure.
The motivations which led the above authors to the model were however
completely different from ours.
One might suspect that a variant of the above model having (statistically)
spherical geometry would lead to a different different universality
class. We checked that this is {\it not} the case by starting with
a central outlet and proceeding as above until the whole domain
is drained.

A typical river obtained by our procedure is shown in
Figure \ref{fig1}.  We have carried out detailed studies of the scaling
properties of the networks.  Our numerical simulations involved
sizes up to $L=192$ with a typical number of different configurations
of the order of $500$.
A summary of our results is presented in
the Table along with the results of observational data\cite{Iri1}.
In order to get a precise estimate of the $\tau$ exponent
we used two different methods. The first consists in
plotting the local slope (which is trivially related to $\tau$)
of the cumulate area distribution (see Fig.\ref{fig2})
\begin{eqnarray} \label{cumulate_area}
P(s,L) &=& \int_{0}^s \; ds^{\prime} \; p(s^{\prime},L)
\end{eqnarray}
as a function of $s$. This is reported in Fig.\ref{fig3}.
An average over all these values yields $\tau=1.406 \pm 0.021$.
On the other hand we performed a finite size analysis of
the exponent extracted at the various sizes (see Fig.\ref{fig4}).
An extrapolation then yields $\tau=1.404 \pm 0.001$.
The value reported in the Table is then the arithmetic
average of these two values. The exponent $\phi$
can be determined by plotting the universal function
as defined in (\ref{pa}). This is also shown in
Fig.\ref{fig5}.
In a similar we computed the exponent $\gamma$
as defined in (\ref{pl}) obtaining
$\gamma=1.612 \pm 0.049$. In Fig. \ref{fig6}
the universal function $f(x)$ is computed
yielding a good collapse for $d_l=1.22$.
The value for $d_l$ can be confirmed by
an independent computation of the typical
length
\begin{eqnarray} \label{moments}
\frac{<l^q>}{<l^{q-1}>} &=& L^{d_l}
\end{eqnarray}
where the averages are referred to the
probability distribution density $\pi(l,L)$.
Using various values of $q$ we found stable
values allowing an estimate as
$d_l=1.22 \pm 0.05$.

The scaling predictions are found to hold very well.
Our exponents agree with those
numerically obtained recently  in Ref. \cite{manna0}
Specifically they reported  $\tau=1.392 \pm 0.010$
and $\gamma=1.628 \pm 0.05$, upon using sizes up
to $L=1024$ but with a less precise data analysis.

This model may be alternately viewed as an optimization
problem that selects the spanning tree that minimizes $\sum _{i} P_{i}$
where $P_{i}$ denotes the erodability of the $i-th$ bond and the sum over
$i$ runs over all the bonds of the tree.\cite{science,bar}
Our model then becomes a special case of the recently introduced optimal
channel networks\cite{OCN1,sun et al} that are constrained to
satisfy a global minimization
of the energy expenditure.\cite{opt}

We note that the resulting network is a union of the invasion 
percolation paths from each of the sites to the outlet\cite{newman et 
al}.  We also note 
that when two paths intersect, they coincide the rest of the way to the 
outlet.   This provides a natural
mechanism for aggregation in our model.  
Also each of the individual 
paths corresponds to the optimal path in a strongly disordered medium 
that was recently shown \cite{Cieplak et al} 
to be characterized by $d_{l} = 1.21 \pm 0.02$.  Note that this value
is somewhat different from the value 1.13 cited by Stark\cite{stark}.  
That value corresponds to the fractal dimension of the {\it geometrical}
shortest path on a percolation cluster and is different from the
fractal dimension of the loopless strands which correspond
to the {\it energetically} shortest path in the sense explained above.

The Hurst 
exponent $H$ obtained\cite{Feder} from a box-counting dimension $d_{l}$ 
would be $2-d_{l}$ and thus in the range 0.77-0.81. With this value of 
$H$ and using the scaling relations in the text, we obtain $\tau = 1.44
\pm 0.04$, $h=0.56\pm 0.01$, and $\gamma = 1.79 \pm 0.02$ in perfect 
agreement with the observational data.  Such a crossover might occur at 
some intermediate length scale beyond which the rivers are no longer
self-similar.

We note that 
the homogeneous analog of our optimal channel network ($P_i$=const) is
the random spanning tree problem, in which all trees
occur with equal probability, $d_l = 5/4$ and the other exponents follow
from our scaling relationships\cite{manna1}.   Our model seems to be in a 
different universality class presumably due to the 
distinct weights associated with the different trees.
The possibility that our result of $d_l \approx 1.21$ may crossover to $5/4$
for much larger sizes cannot of course be ruled out from our data \cite{thanks}

\section{Self-affine river network model}
We now turn to the second model leading to self-affine
($H<1$) river networks.
This corresponds to the homogeneous version of the first model
where the $P_i$'s are all equal. Now the interfacial links all have
an equal probability of being invaded with the usual constraint that loops are
not formed.
This  procedure is akin to the 
well-known Eden growth problem.  A loopless cluster generated in a 
two-dimensional  Eden growth process on a square lattice with a central seed 
site is shown in Figure \ref{fig7}. This model is not new.
Meakin \cite{Meakin} numerically studied the same model although
with different aims. By mapping this problem onto  
a $1+1$ Kardar-Parisi-Zhang equation, Krug and Meakin \cite{KM}
realized that the value $\tau=1.40$ should be exact.
As argued in Refs \cite{cmb},
the individual 
rivers are no longer self-similar but self-affine with a Hurst exponent
of 2/3. The Hurst exponent is larger than the random walk value of
$\frac{1}{2}$  because
the Eden growth process generates strands that compete with each 
other\cite{cmb} and effectively mimic a quenched disordered 
environment\cite{science}.
The exponents predicted from this value of $H$ are shown in the Table.
Figure \ref{fig8} shows the log-log plot of $P(s,L)$ vs. $s$. The data points
analyzed along the same lines as before,
are consistent with the exponent $\tau$ of 1.40 which agrees with the
scaling prediction (see Table). It is remarkable that the self-similar and self-affine
river network models yield very close predictions (indeed $d_l \sim 1.21$
whereas $2/(1+H)=6/5$) for all of the 
scaling exponents even though the underlying mechanisms are very distinct.
We believe this point deserves further
investigation.

\section{Observational data}
In this section we shall review some experimental known
results to provide evidence that the two above
toy models, albeit rather primitive,
are in fact relevant in the interpretation of observational data.
For a more exhaustive analysis see Ref. \cite{rigon et al}.

The network associated with a given natural terrain pertaining to
a river basin can be experimentally analyzed by using the so-called
Digital Elevation Map (DEM) technique (see e.g. \cite{Iri1} and
references therein). In its digitized form, the elevation map 
of a terrain allows the determination of the soil height of areas (pixels)
of order $10^{-2}$ Km$^2$. Thus a fluvial basin is represented
in a objective manner over few (typically 3-4) log scales of linear size.
A flowrate unit is associated with each pixel and the flow contributing
to any pixel follows the steepest descent path through drainage directions.
The resulting network (thus defined by the drainage directions) is
therefore a two-dimensional representation of the three-dimensional landscape.

In Fig. \ref{fig9} a representative network obtained
in this fashion is shown.
This particular network has the exponent $\tau=1.40$ which
is in very good agreement with both the self-similar and
the self-affine models.

It is important to stress that the actual values of the
critical exponents do  vary from basin to basin.
However, within each single river network, all the exponents closely 
satisfy the scaling relationships that we have derived.
The exponent values for the network of Fig. \ref{fig9} is in perfect
agreement with our self-affine (Eden-like) model, although 
other networks often have slightly higher values of $\tau$ \cite{rigon et al}.

The measured values for $d_l$ and $H$ ranges between $1.02$ and
$1.07$ and between $0.75$ to $1.00$ respectively \cite{rigon et al}.
The above models then lie at the outer edge of the observed data in both cases.
We believe that although these two models do represent, at a very schematic level,
two physical mechanisms occurring, at some length scale, in real rivers, our numerical
results show that this is not sufficient and that other effect
possibly combine to yield different numerical exponents.
 
As mentioned earlier, identical statistics can be obtained by using
either self-similar ($d_l>1$, $H=1$) or self-affine ($d_l=1$, $H>1$)
networks, provided that $d_l=\frac{2}{1+H}$. In this respect a direct
measure of the anisotropy of real rivers appears to be crucial. 

Finally we remark that although the observational data
favour  self-affine rivers, the (statistical)
similarity between each basin and its sub-basins, suggests that
river networks are indeed self-similar,
although their actual shape is anisotropic (as expected for
a self-affine network).

\section{Summary}
We have studied two lattice models of spanning trees
that lead naturally to fractal river networks.
We have chosen a self-similar and a self-affine
case on the basis that in nature there are features belonging
to both characters, although it is commonly accepted that
natural rivers are self-affine. The two models have very different
physical motivations.
The self-similar model is directly driven by disorder.
Our choice of heterogeneities is spatially uncorrelated, and any degree
of spatial correlation would decrease the small scale
tortuosity of disorder-dominated paths.
The resulting spanning tree-like structures
exhibit a general similarity to river basins in
overall appearance and very consistent scaling statistics.
Our numerical results are consistent with previous estimates
of the same model introduced on  different physical grounds \cite{manna1}
and corrects some misunderstandings present in the literature \cite{stark}.
More generally, our results show that fractal structures
arise from the minimization of a disorder-dominated total energy functional
and reinforce earlier suggestions \cite{OCN1} on the connections
of optimality with fractal growth.
On the other hand the self-affine model has no disorder in the
definition but it has buried in it a competition mechanism
which {\it effectively} yields quenched disorder.
Again our results are in accord with previous investigations
but have broader consequences in our framework of
analyzing the effects of heterogeneities on the networks.

Two recent investigations are pertinent to our work.
In Ref.\cite{CGMRR} the interplay between quenched disorder
and non-linearity in the landscape evolution was shown
to be relevant for the interpretation of the real rivers field data.
On the other hand, a even more recent renormalization group analysis
of a continuum equation \cite{tadic} suggests that the two disorder-dominated
networks studied here,
the unweighted spanning-trees studied in Ref. \cite{manna0}
and the aforementioned landscape model of Ref. \cite{CGMRR}, all
belong to {\it different} universality classes albeit with
very close exponents. We believe that this scenario is intriguing
and  deserves further attention.

\acknowledgments
We are indebted to  Deepak Dhar, S.S. Manna and Joachim Krug
for useful discussions.
This work was supported by grants from  KBN (grant number 2P302-127),
NASA, NATO
and the Center for Academic Computing at Penn State.

\newpage

\begin{center}
\large
Table
\end{center}

\normalsize

\begin{tabular}{|l|cc|cc|c|}
\hline 
& SELF-SIMILAR  &  &  SELF-AFFINE & & \\
	&Scaling Predictions&Measured& Scaling Predictions&Measured&
River Basins ~\\ 
	&(with $d_{l} = 1.21 \pm 0.02$)	& & (with $H=\frac{2}{3}$) & &~\\
\hline
$d_{l}$	&1.21 $\pm$ 0.02&
1.22 $\pm$ 0.04	& 1& 1 &	1.1 
$\pm$ 0.02 ~\\
$H$	&1&1& $\frac{2}{3}$& $ -- $& 0.75 - 0.80 ~\\	
$\tau$	&1.395 $\pm$ 0.01&	
1.38 $\pm$ 0.03	&$\frac{7}{5}$&1.40$\pm0.02$&	1.43 $\pm$ 0.02 ~\\
$h$	&0.605 $\pm$ 0.01&	
0.62 $\pm$ 0.02	& $\frac{3}{5}$& -- & 0.57 - 0.60 ~\\
$\gamma$ & 1.65 $\pm$ 0.03 & 1.60 $\pm$ 0.05 & $\frac{5}{3}$&-- & 1.8 - 1.9 ~\\	
$d_{F}$	&1.21 $\pm$ 0.02&	
1.21 $\pm$ 0.02 & 1 & 1&	--	~\\
\hline

\end{tabular}

\vskip 3.0cm

The exponents predicted by the scaling arguments, measured in 
our simulations and for river basins.  $d_{F}$ is the fractal dimension 
of the river basin boundary. Note the inconsistency in the observational
data -- $d_{l}$ is greater than 1 suggesting a self-similar network, 
whereas $H<1$ indicating a self-affine structure.

\newpage

\begin{figure}
\caption{Typical self-similar
river network on a $128 \times 128$ lattice obtained by
our optimization procedure. Only  the largest river is shown.
Periodic boundary condition are used
only in the direction transverse to the dominant flow.
The size of the circles is a measure of the value of $s$.}
\label{fig1}
\end{figure}
\begin{figure}
\caption{A log-log plot of the cumulate area distribution $P(s,L)$ vs. $s$,
for lengths ranging from $L=16$ to $L=96$.
Two slopes corresponding to the Scheidegger (dotted line)
and to the Mean Field model (full line) \protect{\cite{takayasu}}
are also shown as a guide for
the eye.}
\label{fig2}
\end{figure}
\begin{figure}
\caption{Effective exponent $\tau$ as computed from the
local slope.}
\label{fig3}
\end{figure}
\begin{figure}
\caption{Plot of the exponent $\tau$ as a function of $1/L$.
The extrapolated value is $\tau=1.404 \pm 0.001$ is indicated
by a star.}
\label{fig4}
\end{figure}
\begin{figure}
\caption{Collapse of the $p(s,L)$ curves for $L$=8 (solid diamonds),
16 (boxes),32 (solid hexagons),64 (stars), and 96 ( closed circles)
according to eq.2,
with $\tau=1.40$ and $\phi=2.0$.}
\label{fig5}
\end{figure}
\begin{figure}
\caption{Plot of the universal function $f(x)$ obtained from
eqn (\protect{\ref{pl}} for the same values as in Fig.5)
with $\gamma=1.61$ and $d_l=1.22$.}
\label{fig6}
\end{figure}
\begin{figure}
\caption{Typical network of self-affine rivers flowing to an outlet in the
center. The network has been obtained by generating an Eden growth process
from a central seed and stopping the growth when the maximal
horizontal distance reached is equal to 64 lattice constants. The size of the
circles is a measure of $s$. Only sites with $s \ge 100$ are shown.}
\label{fig7}
\end{figure}
\begin{figure}
\caption{A log-log plot of the cumulate area distribution $P(s,L)$ vs. $s$ for the
self-affine model. The system sizes are $L$=32, 64, and 256 as indicated in the
figure. The number of samples for the three cases are 1000, 500, and 200
respectively.
The solid line corresponds to the exponent $\tau$=1.40.}
\label{fig8}
\end{figure}
\begin{figure}
\caption{The transportation network of the Johns river drainage basin
in Kentucky, USA.  Its extension is $984 \mbox{Km}^2$.
The measured exponents are: $\tau=1.40$,$H=0.67$.}
\label{fig9}
\end{figure}


\newpage
\begin{references}

\bibitem{Iri1}
D.G. Tarboton, R.L. Bras,
and I. Rodriguez - Iturbe, Water Resour. Res. {\bf 24}, 1317 (1988);
D. Lavall\'ee, S. Lovejoy and D. Schertzer,
{\it Fractals in Geography} (eds. N.S. Lam and L. De Cola), 159
(Prentice Hall, Englewood Cliffs, 1993);
I. Rodriguez-Iturbe, M. Marani, R. Rigon, A. Rinaldo,
Water Resour. Res., {\bf 30}, 3531 (1994);
D.R. Montgomery and W. E. Dietrich, Nature, {\bf 336}, 232 (1988);
Science {\bf 255}, 826 (1992);
S.P. Breyer and R.S. Snow,
Geomorphology {\bf 5}, 143 (1992).

\bibitem{stark} C. P. Stark, Nature {\bf 352}, 423 (1991).

\bibitem{manna0} S. S. Manna and B. Subramanian, Phys. Rev. Lett. {\bf 76},
3460 (1996).

\bibitem{OCN1} I. Rodriguez - Iturbe, A. Rinaldo, R. Rigon, R.L. Bras,
E. Ijjasz - Vasquez, A. Marani,
Water Resour. Res. {\bf 28}, 1095 (1992);
R. Rigon, A. Rinaldo,
I. Rodriguez-Iturbe, E. Ijjasz-Vasquez, R.L. Bras, Water Resour. Res.,
{\bf 29}, 1980 (1993); A. Rinaldo, I. Rodriguez-Iturbe,
R. Rigon, E. Ijjasz-Vasquez and R.L. Bras, Phys. Rev. Lett.,
{\bf 70}, 822 (1993);  for earlier studies linking optimization
principles to drainage networks, see A. D. Howard, Water Res. Res.
{\bf 7}, 863 (1971); ibid  {\bf 26}, 2107 (1990) and references therein.

\bibitem{science} A. Maritan, F. Colaiori, A. Flammini, M. Cieplak and
J. R. Banavar, Science {\bf 272}, 984 (1996); F. Colaiori, A. Flammini,
A. Maritan and J. R. Banavar, Phys. Rev E {\bf 55}, 1298 (1997).

\bibitem{nagel}
Lattice models of river basin evolution are discussed , eg., by
S. Kramer and M. Marder, Phys. Rev. Lett., {\bf 68}, 205 (1992);
R. L. Leheny and S. R. Nagel, Phys. Rev. Lett. {\bf 71},1470 (1993).

\bibitem{sun et al}T. Sun, P. Meakin and T. J{\o}ssang, Phys. Rev. E {\bf
49}, 4865 (1994); ibid {\bf 51}, 5353 (1995); Water Res. Res. {\bf
30}, 2599 (1994); P. Meakin, J. Feder and T. J{\o}ssang, Physica A {\bf
176}, 409 (1991).

\bibitem{Maritan et al} A. Maritan, A. Rinaldo, R. Rigon, A. Giacometti
and I. Rodriguez-Iturbe, Phys. Rev. E {\bf 53}, 1510 (1996). 

\bibitem{Banavar et al} J. R. Banavar, F. Colaiori, A. Flammini, A. Giacometti,
A. Maritan and A. Rinaldo,  Phys. Rev. Lett, {\bf 78}, 4522 (1997)

\bibitem{manna1} S. S. Manna, D. Dhar and S. N. Majumdar, Phys. Rev B {\bf 46},
4471 (1992)

\bibitem{rigon et al} R. Rigon, I. Rodriguez-Iturbe, A. Maritan, A. Giacometti,
D. G. Tarboton and A. Rinaldo, Water Resour. Res. {\bf 32}, 3367 (1996)

\bibitem{hack} J.T. Hack, U.S. Geol. Surv. Prof. Paper {\bf 294}, 1 (1957)

\bibitem{chandler et al}R. Chandler, J. Koplik, Lerman and J. Willemsen,
J. Fluid Mech. {\bf 119}, 249 (1982); R. Lenormand, C. R. Seances, Acad.
Sci. Ser. B {\bf 291}, 279 (1980).

\bibitem{takayasu} See e.g. H. Takayasu, M. Takayasu, A. Provata and G. Huber
J. Stat. Phys. {\bf 65}, 725 (1991).

\bibitem{bar} A.-L. Barabasi, Phys. Rev. Lett. {\bf 76}, 3750 (1996).

\bibitem{opt} An optimal channel network is the spanning tree that
minimizes $\sum_{i} P_i s_i^{\tilde{\gamma}}$ where the sum over $i$
runs over all the bonds of the tree, $P_i$ is the erodability of the $i$-th
bond and $s_i$ is defined in eq.(1). Our model corresponds to a
heterogeneous basin with non uniform $P_i$ and $\tilde{\gamma}$=0.

\bibitem{newman et al} C.M. Newman and D. L. Stein, Phys. Rev. Lett.
{\bf 72}, 2286 (1994).

\bibitem{Cieplak et al} M. Cieplak, A. Maritan and J. R. Banavar,
Phys. Rev. Lett. {\bf 72}, 2320 (1994).

\bibitem{Feder} J. Feder, {\it Fractals} (Plenum, New York, 1988).


\bibitem{thanks} We are grateful to Deepak Dhar for correspondence on
this point.

\bibitem{Meakin} P. Meakin, Phys. Scr. {\bf 45}, 69 (1992); P. Meakin,
J. Phys. A {\bf 20}, L1113 (1987)

\bibitem{KM} J. Krug and P. Meakin, Phys. Rev. A {\bf 40}, 2064 (1989)

\bibitem{cmb} M. Cieplak, A. Maritan, and J. R. Banavar,
Phys. Rev. Lett. {\bf 76}, 3754 (1996).

\bibitem{CGMRR} G. Caldarelli, A. Giacometti, A. Maritan, I. Rodrigues-Iturbe
and A. Rinaldo, Phys. Rev. E {\bf 55}, R4865 (1997); A. Rinaldo, I. Rodrigues-Iturbe,
R. Rigon, E. Ijjazs-Vasques and R. L. Bras, Phys. Rev. Lett. {\bf 70}, 822 (1993).

\bibitem{tadic} B. Tadic, Phys. Rev. Lett. (in press) (1997)


\end{references}
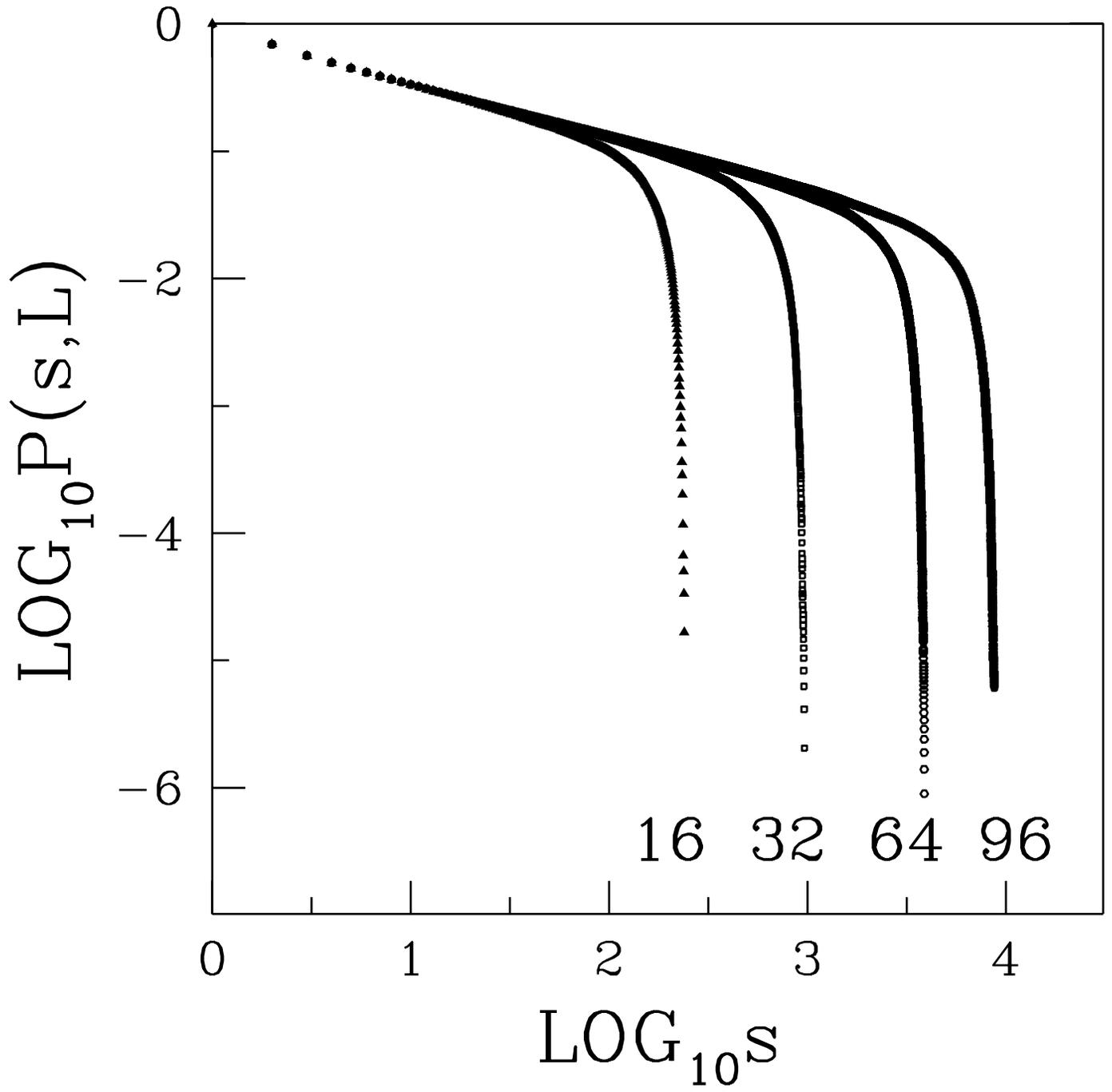
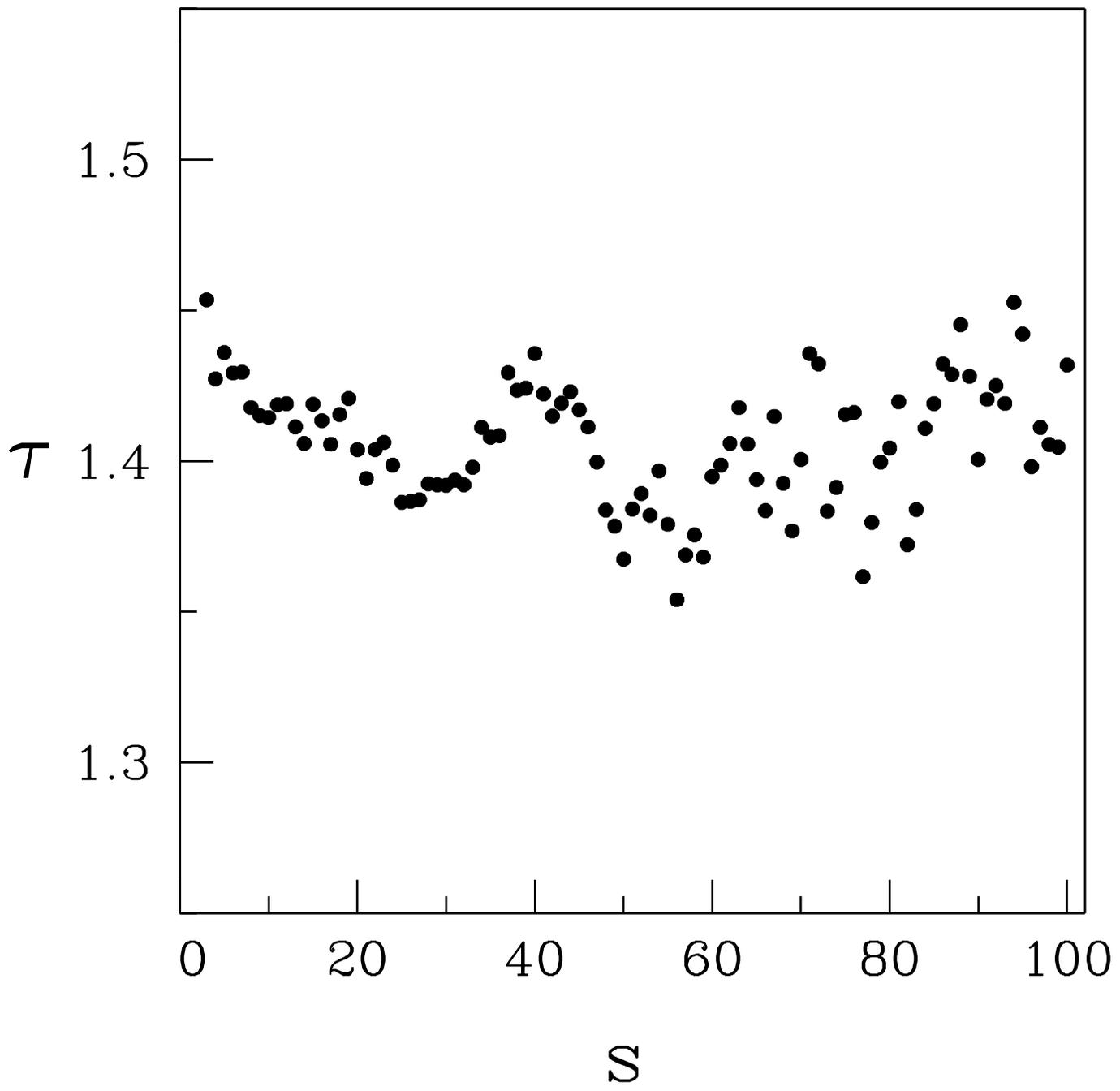
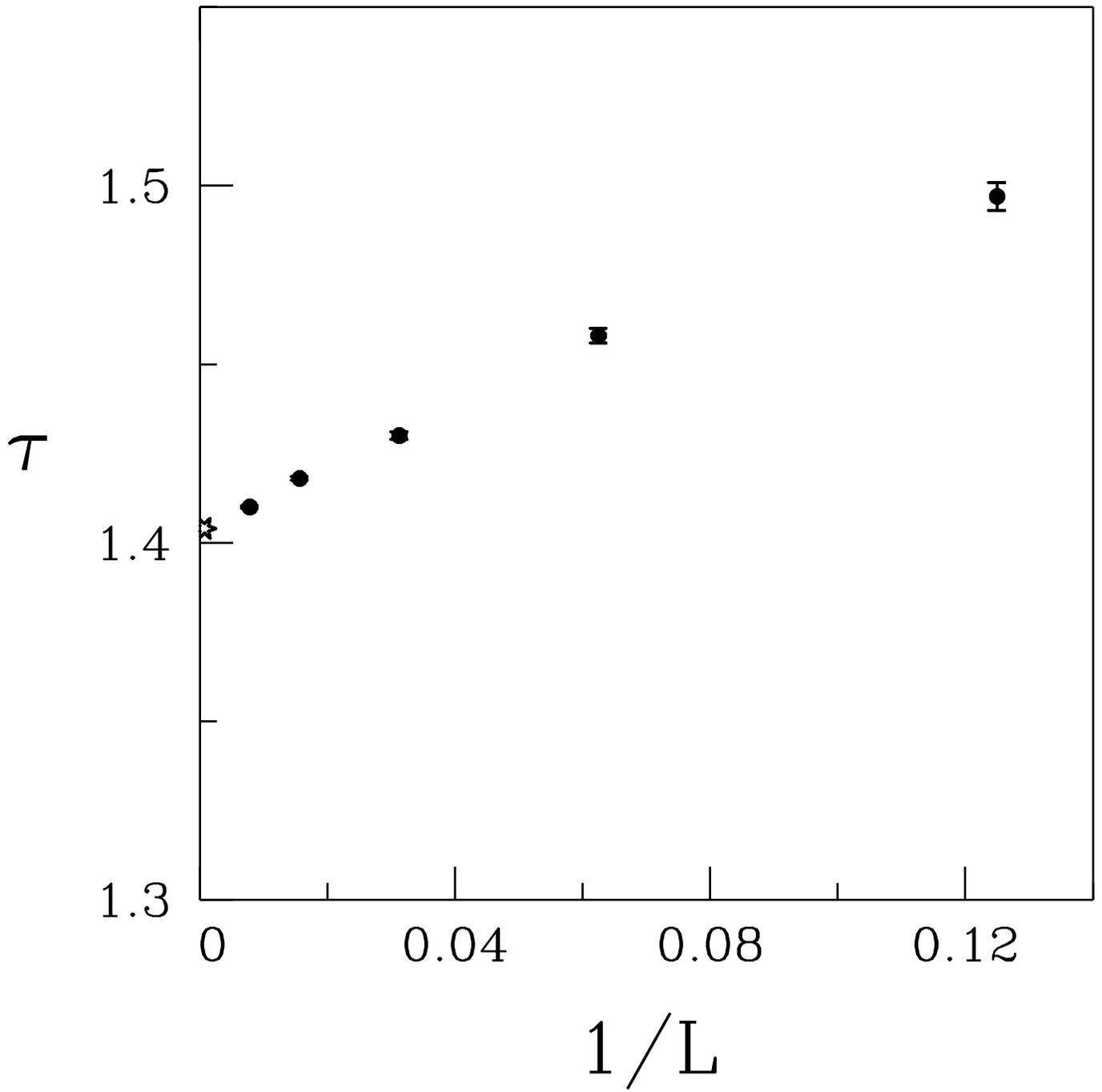
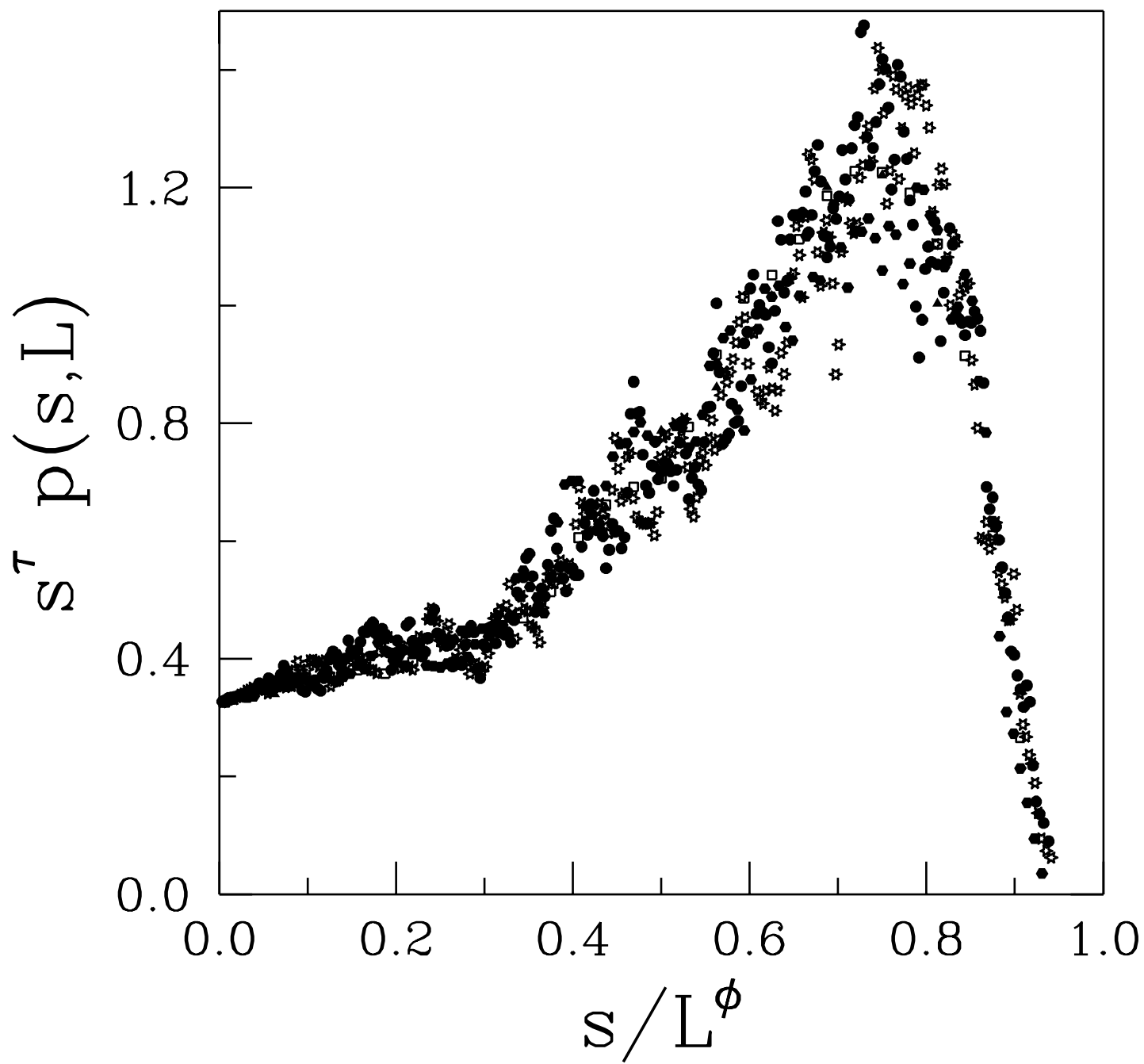
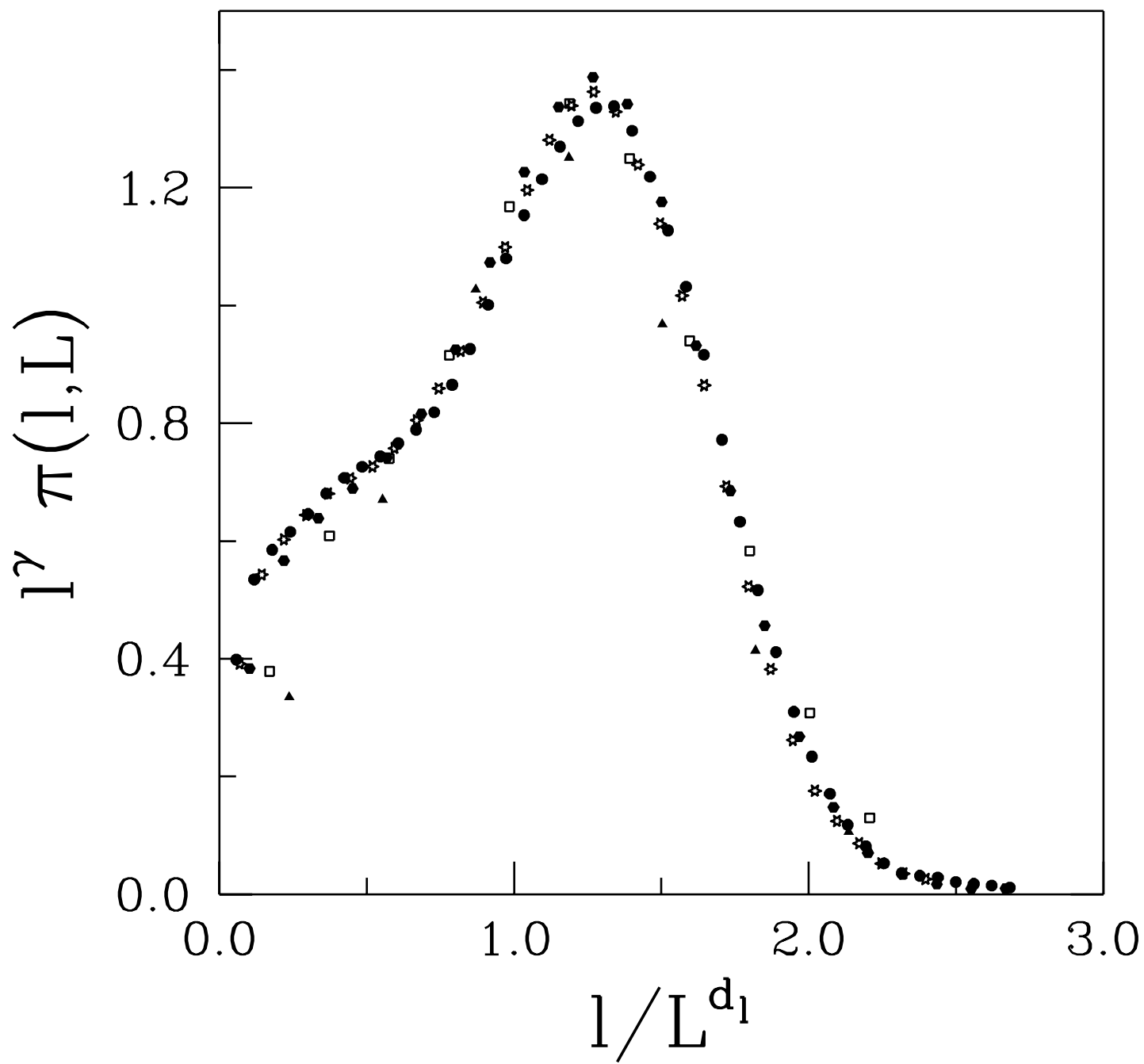
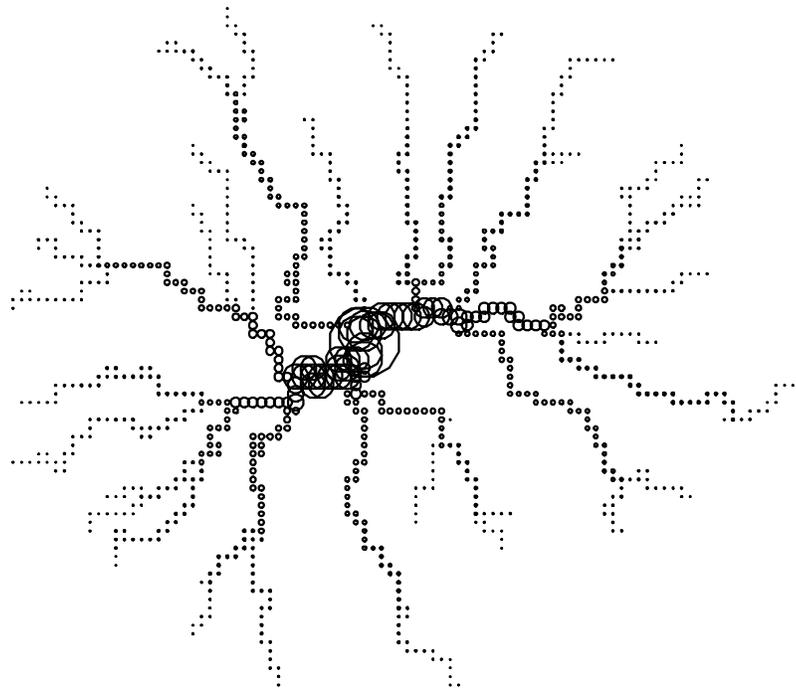
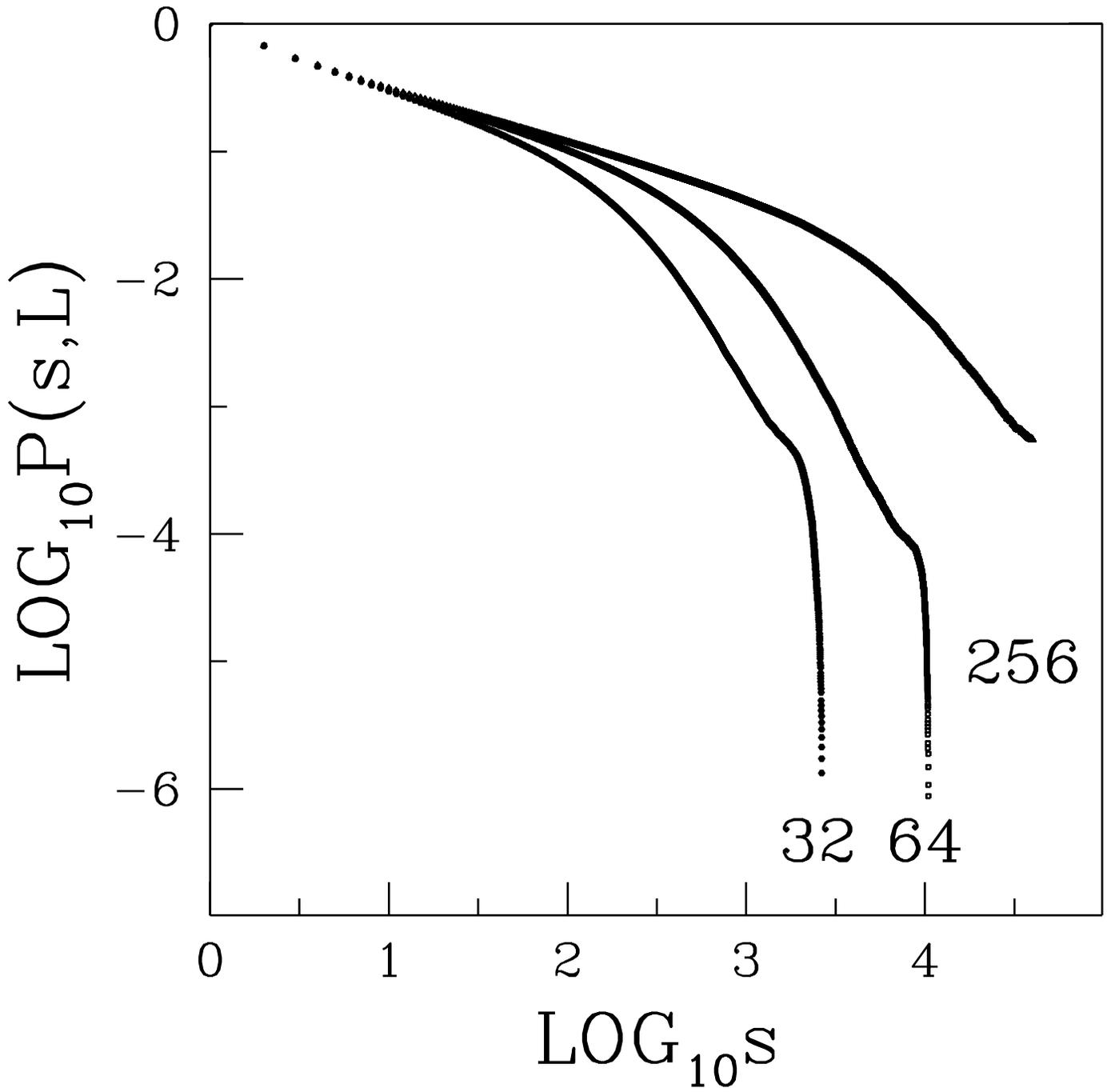
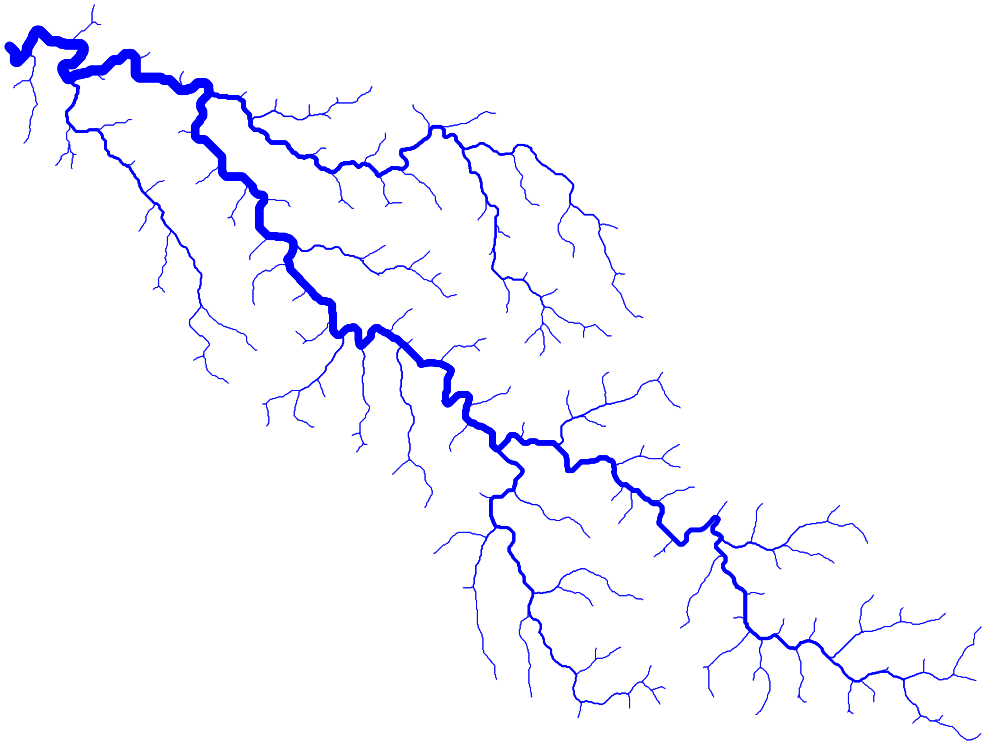
\end{document}